\documentclass[11pt,a4paper]{article}
\usepackage{epsfig}
\usepackage{latexsym}
\begin{document}
\begin{titlepage}
\hspace{11.72cm}MKPH-T-96-3
\begin{center}
\vspace*{1.5cm}
{\bf{\Large{Quasifree pion photoproduction on the deuteron in the \\
  ${\Delta}$ region}}}  \footnote
  {Supported by the Deutsche Forschungsgemeinschaft (SFB 201)} \\
\vspace{0.5cm}
  R.\ Schmidt, H.\ Arenh\"ovel and P.\ Wilhelm\\
  Institut f\"ur Kernphysik           \\
  Johannes Gutenberg -- Universit\"at  \\
  D--55099 Mainz, Germany \\
  March 8, 1996 \\
\end{center}
\vspace{1cm}
\begin{abstract}
\noindent
Photo production of pions on the deuteron is studied in the spectator
nucleon model. The Born terms of the elementary production amplitude
are determined in pseudovector $\pi$N coupling and supplied with a
form factor.  The $\Delta$ resonance is considered both in the $s$ and
the $u$ channel.  The parameters of the $\Delta$ resonance and the
cutoff of the form factors are fixed on the leading photoproduction
multipoles.  Results for total and differential cross sections are
compared with experimental data.  Particular attention is paid to the
role of Pauli correlations of the final state nucleons in the
quasifree case.  The results are compared with those for pion
photoproduction on the nucleon.
\end{abstract}
\end{titlepage}
\renewcommand {\thefootnote} 
              {\arabic{footnote}}
\setcounter{page}{2}

\section{Introduction}
A very interesting topic in medium energy nuclear physics is the
electromagnetic production of mesons on light nuclei in order to study
possible changes of the elementary reaction in a nuclear medium. Of
particular interest is the reaction on the deuteron for the following
reasons.  The first one is that the structure of the deuteron is well
understood in comparison to heavier nuclei. Thus, the low binding
energy of the deuteron allows to compare the contributions of its
constituents to the electromagnetic and hadronic reactions to those
from free nucleons in order to estimate off-shell effects.  Secondly,
the deuteron may be viewed as a neutron target in order to extract
under quasifree conditions the elementary neutron amplitude with the
tacit assumption that binding and final state interaction effects are
small. To assess the validity of this method it is necessary to
estimate the effects of mechanisms beyond this approximation
quantitatively.

First investigations on pion photoproduction on the deuteron go back
to the early fifties, e.g. \cite{ChL51} and \cite{LaF52}, with view on
the general structure of spin flip and no spin flip amplitudes. Later,
a more systematic calculation was done by Laget \cite{Lag77,Lag81} and
Blomqvist and Laget \cite{BlL77}. In this work the influence of pion
rescattering effects and NN final state interaction is approximatly
included within a diagrammatic ansatz. However, they used different
$\Delta$-parametrizations for neutral and charged pion production.

The aim of the present paper is to study pion photoproduction on the
deuteron within the spectator nucleon model using a unified elementary
production operator. We have taken the full on-shell amplitude of the
elementary process but have neglected all kind of final state
interactions and furthermore two-body operators.  In view of the
extraction of neutron data from experiments on the deuteron we put our
emphasis on the special kinematical situation corresponding to the
quasifree pion production on one nucleon. However, this kinematical
condition does not necessarily imply that the pion has been produced
on the active nucleon and not on the spectator.  Therefore, we have
also studied for this kinematical situation the influence of the
second nucleon on the differential cross section due to the
antisymmetrization of the final NN state.

In Sect.\ {\ref{kap2}}, the elementary process of pion photoproduction
on the nucleon is briefly reviewed. The nonresonant amplitudes and the
contribution of the $\Delta$ resonance are given. The model for the
process on the deuteron is outlined in Sect.\ {\ref{kap3}}.  Finally
we present and discuss our results in Sect.\ {\ref{kap4}}.

\setcounter{equation}{0}

\section{Pion photoproduction on the nucleon}
\label{kap2}
We will briefly review pion photoproduction on the nucleon $\gamma (
\vec{k}\, ) + N (\vec{p}\,) \rightarrow N ({\vec{p}}^{\,\prime}\, ) +
\pi (\vec{q}\, )$ because we need the elementary production amplitude
for its implementation into two-nucleon space. In this section we will
first consider the elementary process in the center of mass frame $(
\vec{p} = -\vec{k}\, ,\; \vec{p}^{\,\prime} = -\vec{q}\, ) $ and
generalize later to an arbitrary frame for the calculations on the
deuteron.

The amplitude contains besides nonresonant terms a resonance
contribution from the $\Delta$(1232) excitation which is described in
the framework of an isobar model including its $s$ and $u$ channel
contributions.  For the $\Delta$N$\pi$-vertex we use
\cite{WiA95,WeA78}
\begin{equation}
  v^{\dagger}_{\Delta} = -\frac{i}{m_{\pi}}f_{\Delta N\pi}
  F_{\Delta}(q)
  (-)^{\mu}\tau_{N\Delta,-\mu}\vec{\sigma}_{N\Delta}\!\cdot\!\vec{q}\,
  ,
\end{equation}
where ${\sigma}_{N\Delta}$ and $\tau_{N\Delta}$ are spin and isospin
transition operators, respectively.  We have introduced a hadronic
monopole form factor
\begin{equation}
  F_{\Delta}(q) = \frac{\Lambda^{2}_{\Delta}+q^{2}_{\Delta}}
  {\Lambda^{2}_{\Delta}+q^{2}}\, ,
\end{equation}
where $q_{\Delta}$ is the c.m.\ pion momentum on the top of the
resonance, i.e., when the invariant mass $W_{\pi N}$ of the $\pi$N
state equals the mass of the $\Delta$ resonance
\begin{equation}
  W^{2}_{\pi N} = \left( \sqrt{m_{\pi}^{2} + q_{\Delta}^{2}} +
    \sqrt{M_{N}^{2} + q_{\Delta}^{2}} \right) ^{2} = M_{\Delta}^{2}\,
  .
\end{equation}
For the coupling constant, the cutoff and the $\Delta$ mass we have
used
\begin{equation}
\begin{array}{rcl}
  \frac{f_{\Delta N\pi}^{2}}{4\pi} = 0.31\, ,\qquad & \Lambda_{\Delta}
  = 315 $MeV$\, , & \qquad M_{\Delta} = 1232 $MeV$ \, .
\end{array}
\end{equation}

In the description of the $\gamma N\Delta$ vertex we follow 
\cite{WeA78}. The nonrelativistic limit of the N$\Delta$
current matrix element contains contributions of the magnetic dipole
and electric quadrupole. Since the strength of the electric quadrupole
excitation is much smaller than the magnetic dipole one (see e.\
\hspace{-.06in}g.\ \cite{DaM91}) we will neglect it here.
Then the $\gamma N\Delta$ vertex has the form
\begin{equation}
v_{\gamma N\Delta} = e\frac{G^{M1}_{\Delta N}}{2M_{N}}i\vec{\sigma}_{\Delta
N}\!\cdot\!\vec{k}\times\vec{\epsilon}\,\tau_{\Delta N,0}\, .
\end{equation}
Here $\vec{\epsilon}$ is the photon polarization vector.  The energy
dependent and complex coupling $G^{M1}_{\Delta N}$ is taken from
\cite{WiA95}.  Using the nonrelativistic form of the $\Delta$
propagator, one finally gets for the resonance part of the production
operator
\begin{eqnarray}\label{res}
  t_{\gamma\pi}^{\Delta} & = &
  \hspace{.2cm}
  \frac{F_{\Delta}\left(q \right)}{m_{\pi}}
  \hspace{.1cm}
  \frac{e f_{\Delta N\pi} G^{M1}_{\Delta N}(q)}
  {2\sqrt{E_{p}E_{p^{\prime}}}}
  \hspace{.1cm}
  \frac{\{\tau_{\mu}^{\dagger},\tau_{0}\}-\frac{1}{2}
  [\tau_{\mu}^{\dagger},\tau_{0}] }{3}
  \hspace{.1cm}
  \frac{\vec{\sigma}_{N\Delta}\!\cdot\!
    \vec{q}\, \vec{\sigma}_{\Delta N}\!\cdot\!
    \vec{k}\times\vec{\epsilon}}{W_{\pi N}-M_{\Delta} +\frac{i}{2}
    \Gamma_{\Delta}(W_{\pi N})} \\
  & &
  + \frac{F_{\Delta}\left(0\right)}{m_{\pi}}
  \hspace{.1cm}
  \frac{e f_{\Delta N\pi} G^{M1}_{\Delta N}(0)}
  {2\sqrt{E_{p}E_{p^{\prime}}}}
  \hspace{0.1cm}
  \frac{\{\tau_{\mu}^{\dagger},\tau_{0}\}+
  \frac{1}{2}[\tau_{\mu}^{\dagger},\tau_{0}]}{3}
  \hspace{.1cm}
  \frac{ \vec{\sigma}_{\Delta N}\!\cdot\!
    \vec{k}\times\vec{\epsilon}\, \vec{\sigma}_{N\Delta}\!\cdot\!
    \vec{q}}{E_{\vec{p}^{\, \prime}} - \omega_{\gamma} -
    E^{\Delta}_{\vec{p}^{\, \prime}-\vec{k}} }\, , \nonumber
\end{eqnarray}
where $\Gamma_{\Delta}$ is the energy dependent width of the $\Delta$
resonance above pion threshold
\begin{equation}
  \Gamma_{\Delta}\left( q \right) = \frac{1}{6\pi}\frac{M_{N}}
  {\omega_{\vec{q}}+M_{N}}\frac{q^3}{m_{\pi}^{2}} f^{2}_{\Delta N\pi}
  F^{2}_{\Delta} \left( q \right)\, .
\end{equation}
In the $u$ channel contribution given by the second term in
(\ref{res}), we take the values of the form factor $F_{\Delta}$, the
electromagnetic coupling $G_{\Delta N}^{M1}$ and the width of the
resonance $\Gamma_{\Delta}$ at pion threshold.

For the nonresonant background we used the Born termes in pseudovector
$\pi$N coupling with the coupling constant $\frac{f_{\pi N}^{2}}{4\pi}
= 0.0735$ from \cite{ArZ90}. In order to obtain a better description
of the leading multipoles ($E_{0+}^{1/2}$, $E_{0+}^{3/2}$ and
$M_{1+}^{3/2}$) it was necassary to introduce a form factor
\begin{equation}
  f( q ) = \frac{\Lambda^{2}}{\Lambda^{2} + {q}^{2}}
\end{equation}
with the cutoff $\Lambda = 800$ MeV.

\begin{figure}[ht] 
\vspace{.2in}
\begin{center}
{\epsfig
{figure=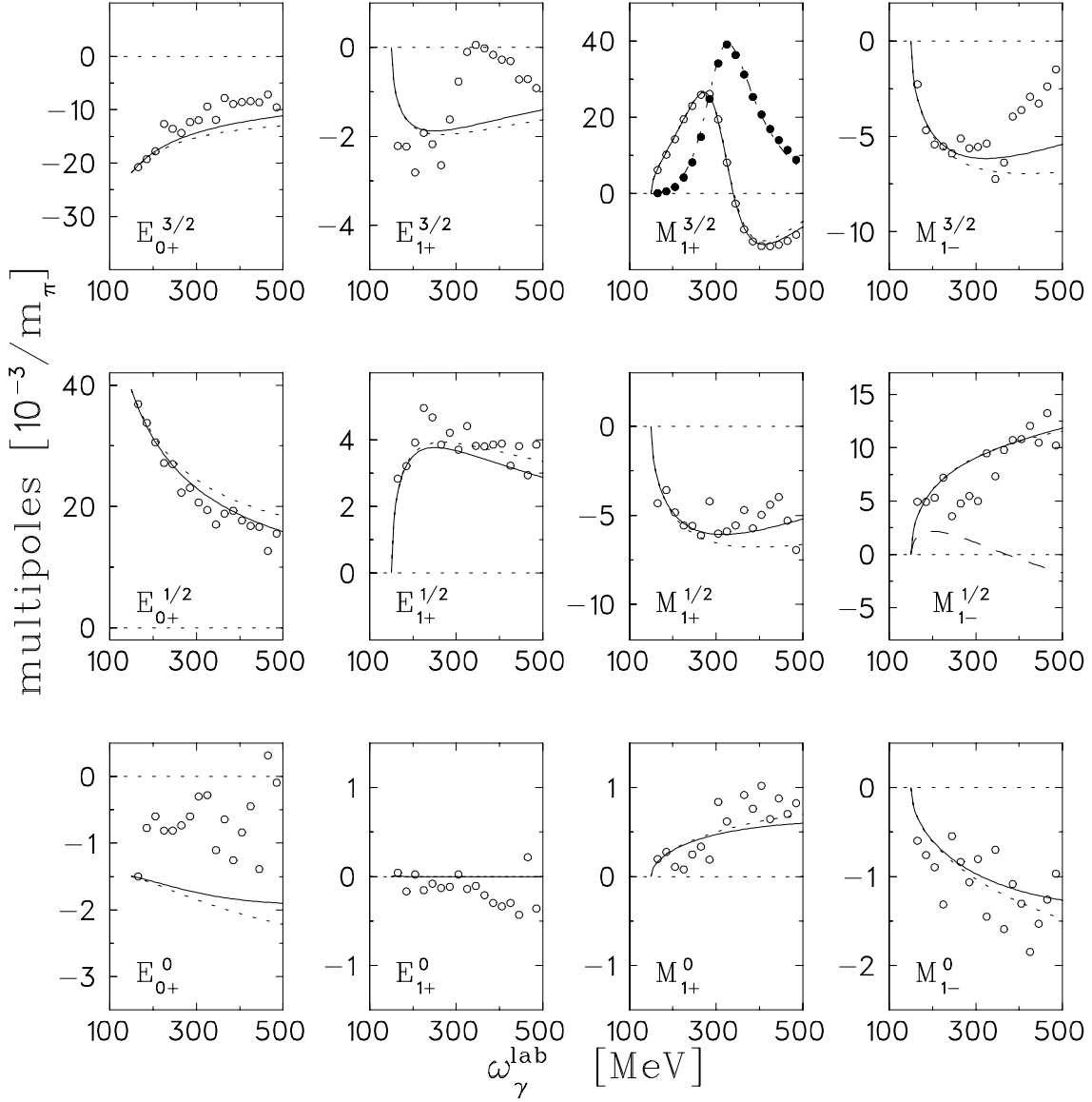,width=15cm}}
\vspace{.1in}
\caption{The $s$- and $p$-wave multipoles in the isospin $3/2$, $1/2$ and $0$
  channels. The full curve shows the real part of the multipoles, the
  dash-dotted one the imaginary part of the $M_{1+}^{3/2}$ for our
  parametrization of the $\Delta$ resonance. The dotted curve shows
  the result for the calculation without form factor in the Born terms
  and for $M_{1-}^{1/2}$ the dashed one when only the $s$ channel of
  the $\Delta$ resonance is taken into account. The data are from
  {\protect\cite{SAID}}. }
\label{fig2}
\end{center}
\end{figure}

The parameters of the $\Delta$ resonance are fixed on the
$M_{1+}^{3/2}$ multipole. Due to the use of a constant $\Delta$ mass
in the $\Delta$ propagator and a different cutoff $\Lambda_{\Delta}$
we had to increase $G^{M1}_{\Delta N}$ from \cite{WiA95} by a factor
of $1.15$ to fit the experimental multipole.  As shown in Fig.\ 
\ref{fig2} the agreement of our results with the data from \cite{SAID}
is very good for both the real and imaginary part of the
$M_{1+}^{3/2}$ multipole up to a photon energy of about 400 MeV.  We
would like to note that for a better description of the real part of
the $M_{1+}^{3/2}$ multipole the suppression of the nonresonant
background by the form factor was essential.  The inclusion of the $u$
channel of the $\Delta$ resonance improved the description of the
$M_{1-}^{1/2}$ multipole.

\begin{figure}[ht] 
\vspace{.2in}
\begin{center}
{\epsfig
{figure=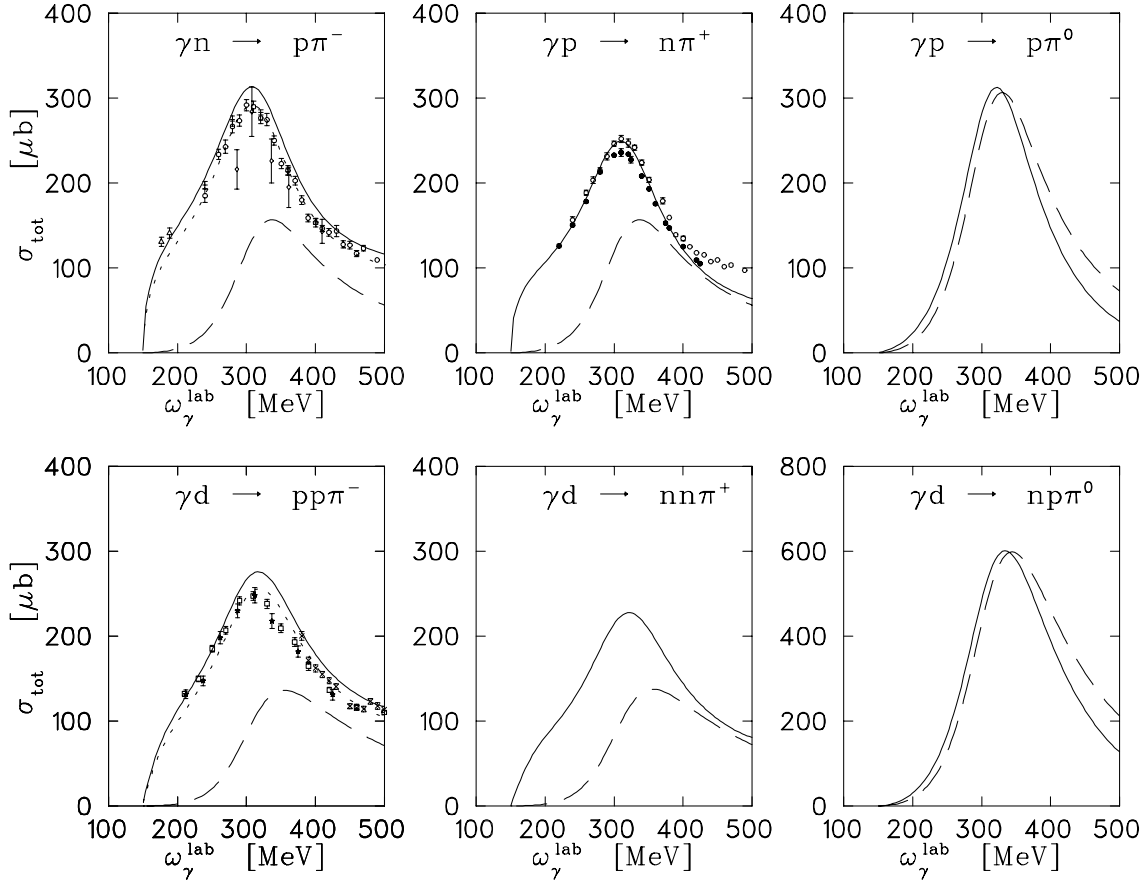,width=15cm}}
\vspace{.1in}
\caption{Total cross sections for pion photoproduction on the nucleon
  and on the deuteron.  The full curve shows the result of the full
  calculation, the long dashed the contribution of the $\Delta$
  resonance.  The dotted curve shows the full result using the
  rescaled nonresonant background.  The experimental data are from
  {\protect\cite{CoB75}} ($\Diamond$), {\protect\cite{SaM84}}
  ($\triangle$), {\protect\cite{HoK74}} ($+$), {\protect\cite{FuK77}}
  ($ \circ $) and {\protect\cite{FiH72}} ($\bullet$) for the reaction
  on the nucleon and from {\protect\cite{BeB73}} ($\Box$),
  {\protect\cite{ChD75}} ($\star$), and {\protect\cite{AsE90}}
  ($\times$) for pion photoproduction on the deuteron. }
\label{fig3}
\end{center}
\end{figure}

In Fig.\ \ref{fig2} the results for the other leading multipoles in
the three isospin channels are also shown. The description of these
multipoles is in general quite satisfactory. Due to the neglect of the
$E_{1+}$ excitation of the $\Delta$ resonance we fail to describe the
behaviour of the $E_{1+}^{3/2}$ multipole. Furthermore the imaginary
part of all multipoles apart from the $M_{1+}^{3/2}$ vanishes in our
model because we have not included other resonances.  In the upper
part of Fig.\ \ref{fig3} we show the total cross sections for pion
photoproduction on the nucleon compared to experimental data. In
contrast to \cite{BlL77} we used in our calculation the same
parametrization of the $\Delta$ resonance both for neutral and charged
pion photoproduction. The price we have to pay in order to achieve a
good description of all physical channels is the introduction of the
above mentioned form factor into the nonresonant background amplitude.
The agreement with the experimental data is good up to an invariant
mass of about 1300 MeV ($\omega^{lab} = 430$Mev) in the case of
$\pi^{+}$ production. The $\pi^{-}$ data are slightly overestimated in
the resonance region by our calculation.

\section{Pion photoproduction on the deuteron}
\label{kap3}
It is reasonable to expect that the dominant process will be the
quasifree reaction on one nucleon while the other acts merely as a
spectator remaining at rest in the laboratory system so that the final
state interaction between the nucleons and pion rescattering on the
spectator nucleon may safely be neglected.  In this spectator nucleon
ansatz we will use the elementary amplitude discussed in Sect.\ 
\ref{kap2}. Due to the Fermi motion in the initial state we have to
use this amplitude in a form which is not restricted to a special
frame of reference.  For the nonresonant background this is achieved
by evaluation of the corresponding Feynman diagrams in an arbitrary
frame. This leads to the following expression for the nonresonant
production operator
\begin{eqnarray}
  t_{\gamma\pi}^{nonres} & = & \frac{if_{\pi N}f(q_{c.m.})}{m_{\pi}}
{\Bigg\{ }
\left[
\vec{\sigma}\!\cdot\!\vec{\epsilon}
+ \frac{\vec{q}\!\cdot\!\vec{\epsilon}
\vec{\sigma}\!\cdot\!(\vec{q}-\vec{k})}{\omega_{\vec{q}-\vec{k}}}
\left(\frac{1}{\omega_{\vec{q}}-\omega_{\vec{q}-\vec{k}}-
\omega_{\gamma}}+\frac{1}{\omega_{\gamma}-\omega_{\vec{q}-\vec{k}}-
 \omega_{\vec{q}}}\right)\right]
\, [\,\hat{e},\tau_{\mu}^{+}\, ]
\nonumber \\
& &
-\frac{\tau_{\mu}^{+}\vec{\sigma}\!\cdot\!\vec{q}\left( 2\left(
\vec{p}^{\, \prime}+\vec{q}\right)\!\cdot\!\vec{\epsilon}
\,\hat{e}+i\vec{\sigma}
\!\cdot\!\vec{k}\times\vec{\epsilon}\left(\hat{e}+\hat{\kappa}\right)\right)}
{2E_{\vec{p}^{\,\prime}+\vec{q}}\left(\omega_{\vec{q}} +
E_{\vec{p}^{\, \prime}}-E_{\vec{p}^{\,\prime}+\vec{q}}
\right)}
-\frac{\left(2\vec{p}^{\,\prime}\!\cdot\!\vec{\epsilon}\,\hat{e}
+i\vec{\sigma}\!\cdot\!\vec{k}\times\vec{\epsilon}
\left(\hat{e}+\hat{\kappa}\right)\right)\tau_{\mu}^{+}
\vec{\sigma}\!\cdot\!\vec{q}}
{2E_{\vec{p}^{\,\prime}-\vec{k}}
\left(E_{\vec{p}^{\,\prime}}-E_{\vec{p}^{\,\prime}-\vec{k}}-\omega_{\gamma}
\right)} \nonumber \\
& &
\left.
+M_{N} \omega_{\vec{q}} \vec{\sigma}\!\cdot\!\vec{\epsilon}
\left(
\frac{\tau_{\mu}^{+}\hat{e}}{E_{\vec{p}^{\,\prime}+\vec{q}}\left(
E_{\vec{p}^{\,\prime}+\vec{q}}+E_{\vec{p}^{\,\prime}}+\omega_{\vec{q}}
\right)}
+
\frac{\hat{e}\tau_{\mu}^{+}}{E_{\vec{p}^{\,\prime}-\vec{k}}\left(
E_{\vec{p}^{\,\prime}-\vec{k}}+E_{\vec{p}^{\,\prime}}-\omega_{\gamma}
\right)}
\right)
\right\}\, ,
\end{eqnarray}
where $\hat{e}$ and $\hat{\kappa}$ denote nucleon charge and anomalous
magnetic momentum.  In the resonance amplitude
$t_{\gamma\pi}^{\Delta}$ in (\ref{res}), the photon and pion momenta
have to be replaced by the relative photon-nucleon momentum
\begin{equation}
  \vec{k}_{\gamma N} = \frac{M_{N}\vec{k} - \left(
      M_{\Delta}-M_{N}\right)\,  \vec{p}}{M_{\Delta}}
\end{equation}
and respective pion-nucleon momentum
\begin{equation}
  \vec{q}_{\pi N}=\frac{M_{N}\vec{q} -
    \omega_{\vec{q}}\vec{p}^{\,\prime}} {M_{N}+\omega_{\vec{q}}}\, .
\end{equation}
In the expressions of the form factors, the coupling
$G^{M1}_{\Delta N}$ and the width of the $\Delta$ resonance we use the
c.m. pion momentum as given by the
invariant mass of the $\pi$N subsystem
\begin{eqnarray}
  W_{\pi N}^{2} & = & \left( \sqrt{M_{N}^{2} + \vec{p}^{\,\prime\; 2}}
    + \sqrt{m_{\pi}^{2} + \vec{q}^{\, 2}} \right)^{2} - \left(
    \vec{p}^{\,\prime} + \vec{q} \right)^{2} \nonumber \\ & = &
  \left(\sqrt{M_{N}^{2} + \vec{q}_{c.m.}^{\, 2}} + \sqrt{m_{\pi}^{2} +
      \vec{q}_{c.m.}^{\, 2}} \right)^{2}
\end{eqnarray}
With $t_{\gamma\pi} = t^{\Delta}_{\gamma\pi} + t^{nonres}_{\gamma\pi}$
we use as pion photoproduction operator in the two-nucleon system
\begin{equation}\label{g17}
  t_{\gamma\pi}^{NN}\left( 1,2 \right) = t_{\gamma\pi} \left( 1
  \right) + t_{\gamma\pi} \left( 2 \right)\, .
\end{equation}

Since the final state interaction is neglected we take plane waves for
the pion and the NN-final states. For the spin $( |s m_{s}\rangle )$
and isospin $( |t m_{t}\rangle )$ part of the two nucleon wave
functions we use a coupled basis.  The complete antisymmetric final NN
wave functions is
\begin{equation}
  |\vec{p}_{1},\vec{p}_{2},s m_{s},t m_{t} \rangle =
  \frac{1}{\sqrt{2}}\left(
    |\vec{p}_{1}\rangle^{(1)}|\vec{p}_{2}\rangle^{(2)} + (-)^{1+s+t}
    |\vec{p}_{2}\rangle^{(1)}|\vec{p}_{1}\rangle^{(2)}\right)|s
  m_{s}\rangle |t m_{t}\rangle\, .
\end{equation}
Only the $t = 1$ channel contributes in the case of charged pions
whereas for $\pi^{0}$ production both $t = 0$ and $t = 1$ channels
have to be taken into account.  Now one easily writes down the matrix
element for pion photoproduction on the deuteron
\begin{eqnarray}
  \langle \vec{p}_{1},\vec{p}_{2},s m_{s},t m_{t}
  |t_{\gamma\pi}^{NN}|\vec{d},m_{d} \rangle & = & \frac{1}{2} \int
  \frac{d^{3}p^{\prime}_{1}}{(2\pi)^{3}} \int
  \frac{d^{3}p^{\prime}_{2}}{(2\pi)^{3}}
  \frac{M_{N}}{E_{\vec{p}^{\,\prime}_{1}}}
  \frac{M_{N}}{E_{\vec{p}^{\,\prime}_{2}}} \\ & & \times
  \sum_{m_{s}^{\prime}}\, \langle \,\vec{p}_{1}\vec{p}_{2},s m_{s},t
  m_{t} |\, t_{\gamma\pi}^{NN}
  |\,\vec{p}_{1}^{\,\prime}\vec{p}_{2}^{\,\prime}, 1 m_{s}^{\prime}, 0
  0 \rangle \langle\,\vec{p}_{1}^{\,\prime}\vec{p}_{2}^{\,\prime}, 1
  m_{s}^{\prime}|\,\vec{d}, m_{d}\rangle \, . \nonumber
\end{eqnarray}
with $t_{\gamma\pi}^{NN}$ from (\ref{g17}), where we have used the covariant
normalization
\begin{eqnarray}
\langle \vec{p}^{\,\prime} |\, \vec{p}\,\rangle & = &
(2\pi)^{3}\frac{E_{p}}{M_{N}}
\,\delta^{3}\!\left( \vec{p}^{\,\prime}-\vec{p}\,\right)\; , \\
\langle \vec{d}^{\,\prime} |\, \vec{d}\,\rangle & = &
(2\pi)^{3} 2E_{d}
\,\delta^{3}\!\left( \vec{d}^{\,\prime}-\vec{d}\,\right)\, ,
\end{eqnarray}
and the deuteron wave function in the form
\begin{equation}
  \langle\,\vec{p}_{1}^{\,\prime}\vec{p}_{2}^{\,\prime}, 1
  m_{s}^{\prime}|\,\vec{d}, m_{d}\rangle = (2\pi)^{3}
  \delta^{3}\left(\,
    \vec{d}-\vec{p}_{1}^{\,\prime}-\vec{p}_{2}^{\,\prime} \,\right)
  \frac{\sqrt{E_{\vec{p}_{1}^{\,\prime}}E_{\vec{p}_{2}^{\,\prime}}}}
  {M_{N}} \tilde{\Psi}_{m_{s}^{\prime},m_{d}}\left(\,
    \frac{1}{2}\left( \vec{p}_{1}^{\,\prime}-\vec{p}_{2}^{\,\prime}\,
    \right)\right)
\end{equation}
with
\begin{equation}
  \tilde{\Psi}_{m_{s},m_{d}}(\vec{p}) =
  (2\pi)^{\frac{3}{2}}\sqrt{2E_{d}}
  \sum_{L=0,2}\sum_{m_{L}}i^{L}u_{L}(p)Y_{Lm_{L}}(\hat{p})
  (Lm_{L}1m_{s}|1m_{d})\, .
\end{equation}
In the laboratory frame (deuteron rest frame) one finds for the matrix
element the following expression
\begin{eqnarray}\label{g16}
  \langle \vec{p}_{1},\vec{p}_{2},s m_{s},t m_{t} |t_{\gamma\pi}^{NN}|
  \vec{d}=0,m_{d} \rangle &=& \sqrt{2}\sum_{m_{s}^{\prime}}\langle s
  m_{s}| \langle t m_{t}|\,\Big( \langle
  \vec{p}_{1}|t_{\gamma\pi}^{(1)}|-\vec{p}_{2}\rangle
  \tilde{\Psi}_{m_{s}^{\prime},m_{d}}(\vec{p}_{2}) \\ & &
  \hspace{.56in} +(-)^{1+s+t} \langle
  \vec{p}_{2}|t_{\gamma\pi}^{(1)}|-\vec{p}_{1}\rangle
  \tilde{\Psi}_{m_{s}^{\prime},m_{d}}(\vec{p}_{1})\Big)\,|1
  m_{s}^{\prime} \rangle |00\rangle \, . \nonumber
\end{eqnarray}
Note that in (\ref{g16}) all spin and isospin operators act on nucleon
``1''.

The differential cross section is then given by
\begin{equation}\label{g18}
  d\sigma = (2\pi)^{-5}\delta^{4}\left( k+d-p_{1}-p_{2}-q\right)
  \frac{1}{|\vec{v}_{\gamma}\!-\!\vec{v}_{d}|}
  \frac{d^{3}q}{2\omega_{\vec{q}}} \frac{d^{3}p_{1}}{2E_{\vec{p}_{1}}}
  \frac{d^{3}p_{2}}{2E_{\vec{p}_{2}}}
  \frac{M_{N}^{2}}{2\omega_{\gamma}E_{d}} \frac{1}{6}\sum_{m_{\gamma}
    ,m_{d}} \sum_{s,m_{s}} \sum_{t,m_{t}}
|t^{NN}_{\gamma\pi}|^{2} . 
\end{equation}
We would like to mention that by decoupling the spin wave functions in
(\ref{g16}) and performing the spin summation in (\ref{g18}) explicitly,
we formally reproduce the expression for
the differential cross section given in \cite{BlL77} for an uncoupled spin
basis.

\section{Results and discussions}
\label{kap4}
Now we will present our results on pion photoproduction on the
deuteron using the deuteron wave function of the Bonn potential (full
model) \cite{MaH87}.  The discussion is diveded into two parts. In the
first we compare with experimental data of the total cross section
$\sigma_{tot}$ and the semi-exclusive differential cross section
$d\sigma /d\Omega_{\pi}$ as a test of our model. It is obtained from
the fully exclusive cross section $d^{5}\sigma
/d\Omega_{\pi}dq_{\pi}d\Omega_{\vec{P}}$ where $\vec{P}$ is the total
momentum of the two nucleons in the final state.  In order to include
symmetrically the contributions of both nucleons in the numerical
integration, we have chosen $\Omega_{\vec{P}}$ to be an independent
quantity.  In the second part, we have studied the influence of Pauli
correlations in the final state on the differential cross section in a
special kinematical situation which allows a direct comparison to the
free nucleon process.

\subsection{Comparison with experimental data}
\label{kap4_1}
We will concentrate our discussion on $\pi^{-}$ photoproduction, since
no data for $\pi^{+}$ and $\pi^{0}$ production in the $\Delta$ region
are available. In the lower part of Fig.\ \ref{fig3} we show the total
cross sections for pion photoproduction on the deuteron which can be
compared to the free nucleon case in the upper part.  In the case of
$\pi^{-}$ production on the neutron we compare our results with data
from the inverse reaction $\pi^{-}n \rightarrow p\gamma$
\cite{CoB75,SaM84} and with data extracted from experiments on the
deuteron \cite{HoK74,FuK77}. For the pion production operator of
Sect.\ \ref{kap2} we find an overestimation of the total cross section
in the resonance peak for $\pi^{-}$ production both on the neutron and
on the deuteron. For this reason we have calculated the cross sections
using a nonresonant background for a lower $\pi$N coupling constant
$f^{2}_{\pi N}/4\pi = 0.069$. In this case the description of the
experimental data in the resonance peak is good.  However, for photon
energies below the resonance up to about 280 MeV the theoretical
results underestimate the data significantly.

\begin{figure}[ht]
\vspace{.2in}
\begin{center}
{\epsfig
{figure=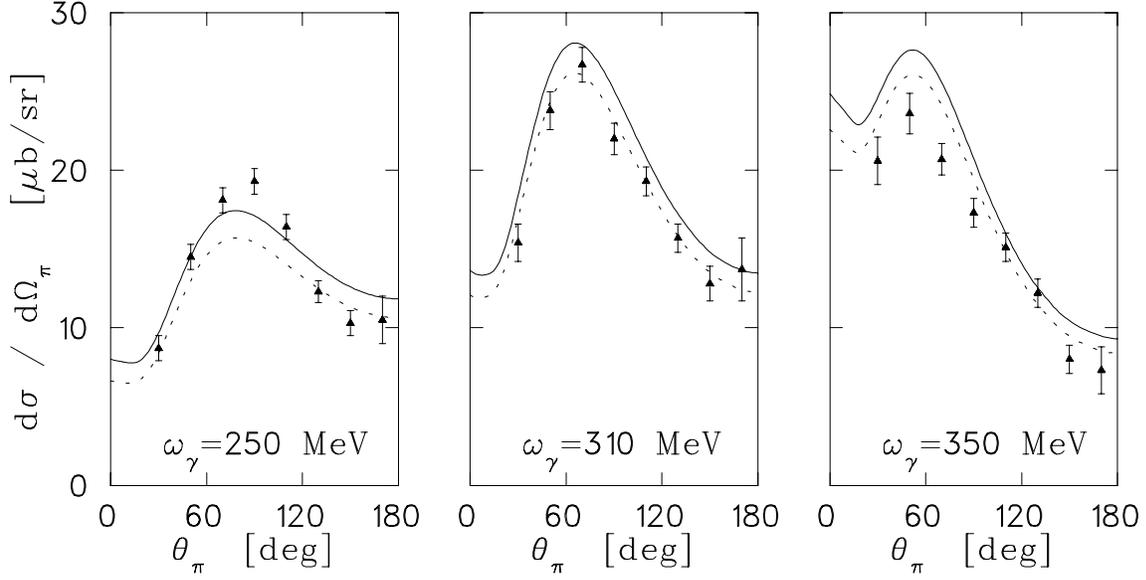,width=15cm}}
\vspace{.1in}
\caption{Differential cross section $d\sigma /d\Omega_{\pi}$ for
  $\gamma d \longrightarrow pp\pi^{-}$. See caption of Fig.\ 
  {\protect\ref{fig3}} for the meaning of the curves.  The
  experimental data are from {\protect\cite{BeB73}}.}
\label{fig7}
\end{center}
\end{figure}

The differential cross sections $d\sigma /d\Omega_{\pi}$ are shown in
Fig.\ \ref{fig7}.  
The use of the pion production operator of Sect.\ 
\ref{kap2} leads to an overestimation of the data for energies above
300 MeV. However, the shape of the curves is in good agreement with the
experiment. Using the lower $\pi$N coupling $f^{2}_{\pi N}/4\pi = 0.069$
we found a good description of the data for photon energies above 300 MeV.

\subsection{Quasifree kinematics}
\label{subsecquas}
Now we will discuss the effect of the Pauli correlations, i.e.,
antisymmetrization of the final NN state for quasifree kinematics.  In
order to compare with pion photoproduction on the nucleon we did the
calculation in the rest frame of the quasifree $\pi$N subsystem (see
Fig.\ \ref{fig10}a) so that its contribution to the t-matrix
will be the same as that of a free nucleon in the
center of mass frame apart from some minor approximations which are
discussed in Appendix \ref{app3}.  For the genuine quasifree case in
Fig.\ \ref{fig10}a the relative momentum of the two nucleons in the
initial state vanishes while for the spectator contribution in Fig.\ 
\ref{fig10}b, i.e., reaction on the spectator, the relative momentum
is $\vec{q}-\vec{k}$. In particular for pion production at backward
angles, the contribution of the spectator nucleon will be largely
suppressed by the deuteron wave function because of the large momentum
mismatch, but this will not work for small pion emission angles.

In Appendix \ref{app3} we show that the neglect of the contribution of
the nucleon with momentum $-\vec{k}$ in the final state leads to a
very simple relation between the differential cross sections of the
quasifree and the elementary processes, namely
\begin{equation}\label{g19}
  {\frac{d^{5}\sigma^{\, q.f.}}{d^{3}P_{\pi N} d\Omega_{\pi}}}
  =\frac{u_{0}^{2}(0)}{4\pi} {\frac{d\sigma^{\,
        el.}}{d\Omega_{\pi}}}\, ,
\end{equation}
where $u_{0}(0)$ is the momentum space wave function of the s-wave at
zero momentum and $\vec{P}_{\pi N}$ is the total momentum of the
$\pi$N system.  The results for the ratio of these cross sections are
shown in Fig.\ \ref{fig11}.  For backward pion emission the relation
(\ref{g19}) is verified within about $1 \%$ justifying the neglect of
the second nucleon for larger pion angles.  However, for forward pion
emission this is not any more the case.  Here one finds a decrease of
the cross section up to about $10\%$ for small pion angles
due to Pauli correlations. This means that even for 
quasifree kinematics one will get a modification of the observables
for certain kinematics
which is not negligible. In our opinion, this is a very important point
for the extraction of neutron data from experiments on the deuteron.

\section{Summary}
We have investigated pion photoproduction on the deuteron in the
$\Delta (1232)$ resonance region in the spectator nucleon model
neglecting final state interactions and two body processes.  Within
this framework, the t-matrix is given as a linear combination of the
on-shell matrix elements of pion photoproduction on the two nucleons.
The elementary amplitude is fitted to the pion photo production
multipoles. We have presented results for total and differential
cross sections. A comparison with experimental data has been possible
only for $\pi^{-}$ production, where we found a slight overestimation
of the data which largely go back to an overestimation of the elementary
reaction on the neutron.

Particular attention was paid to the quasifree case for which we have
studied the effect of the antisymmetrisation of the NN-final state.
Even for quasifree kinematics where the process is dominated by the
reaction on one of the nucleons, we found an effect of about $10\% $
in the differential cross section due to presence of the second
nucleon.

The differences between the theoretical results and the experimental
data show very clearly that the calculation of pion photoproduction on
nuclei in the nucleon spectator model can only be considered as a
first step towards a more realistic description of this process. The
studies discussed here will serve as the basis for further
investigations including the dynamics of the $\pi$NN system in a more
satisfactory way.

\newpage
\appendix
\renewcommand{\thesection}{}
\renewcommand{\theequation}{\Alph{section}.\arabic{equation}}  
\setcounter{equation}{0}

\section{\hspace{-.55cm}Appendix \hspace{.15cm} Evaluation of Equation 
({\protect\ref{g19}})}
\label{app3}
\begin{figure}[ht] 
\vspace{.2in}
\begin{center}
{\epsfig
{figure=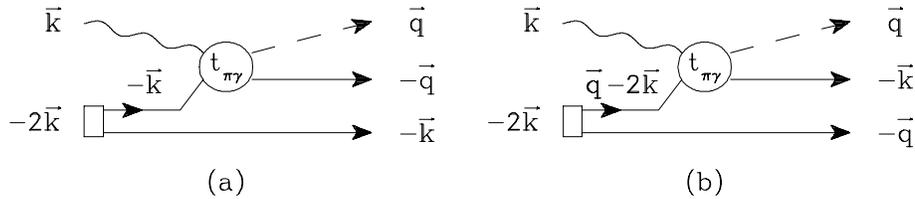,width=12cm}}
\vspace{.1in}
\caption{The contributions of the two nucleons in quasifree kinematics:
(a) quasifree case, (b) spectator contribution.}
\label{fig10}
\end{center}
\end{figure}

\begin{figure}[ht] 
\vspace{.2in}
\begin{center}
{\epsfig
{figure=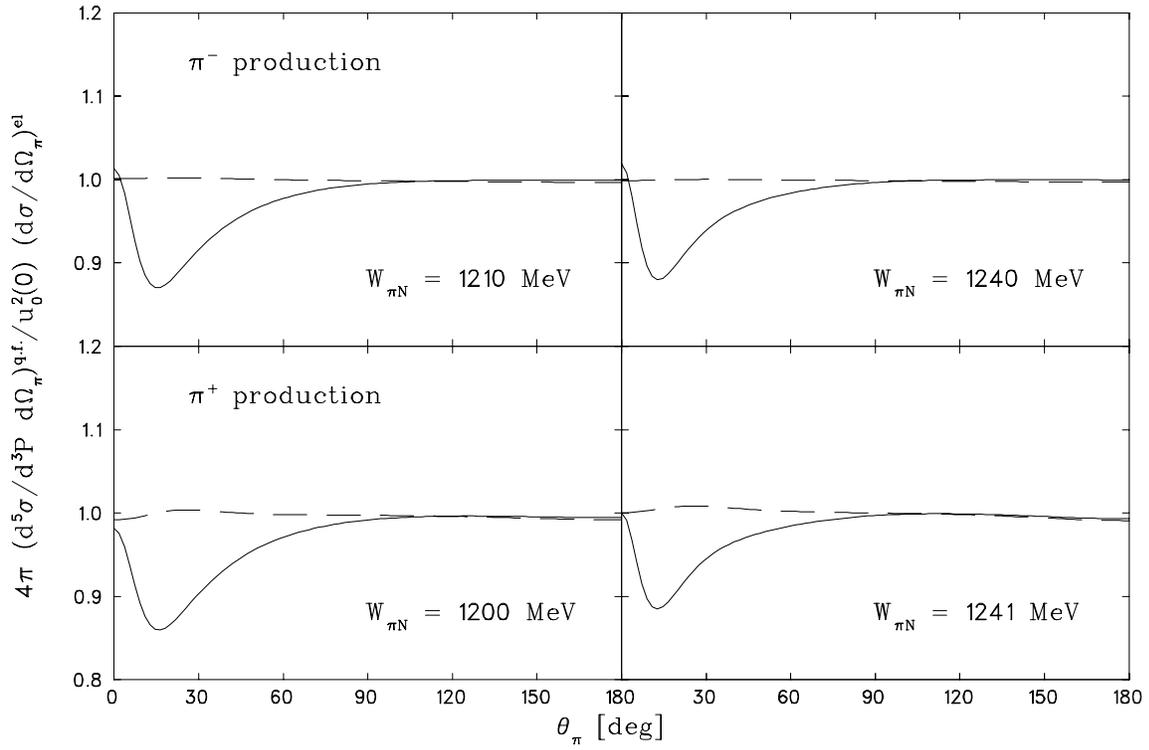,width=15cm}}
\vspace{.1in}
\caption{The ratio of the differential cross sections of $\gamma d
  \longrightarrow NN\pi$ and $\gamma N \longrightarrow N\pi$. The full
  curve shows the result for the full calculation on the deuteron, the
  dashed curve only the contribution of the nucleon which is
  distinguished by the quasifree kinematic (Fig.\ {\protect\ref{fig10}}a).}
\label{fig11}
\end{center}
\end{figure}

Here we will give a brief derivation of equation (\ref{g19}) and a
discussion of the approximations used. The general expression for the
exclusive differential cross section is given in (\ref{g18}).
Evaluating the $\delta$-functions with respect to the independent
kinematical quantities $\vec{P}_{\pi N}$ and $\Omega_{\pi}$, one finds
in the frame of reference defined in Fig.\ \ref{fig10}
\begin{equation}\label{c1}
  \frac{d^{5}\sigma}{d^{3}P_{\pi N}d\Omega_{\pi}} =
  \frac{1}{(2\pi)^{5}}
  \frac{1}{(E_{d}+2\omega_{\gamma})(E_{\vec{q}}+\omega_{\vec{q}})}
  \frac{|\vec{q}\, |}{\omega_{\gamma}}\frac{M^{2}_{N}}
  {E_{\vec{k}}}\frac{1}{8} \frac{1}{6}\sum_{m_{\gamma} ,m_{d}}
  \sum_{s,m_{s}} |t^{NN}_{\pi\gamma}|^{2} \, .
\end{equation}
Taking into account only the contribution of the nucleon with momentum
$-\vec{q}$ in the final state the corresponding t-matrix
$\tilde{t}_{\pi\gamma}^{NN}$ is given by
\begin{eqnarray}
  \langle -\vec{q},-\vec{k},s m_{s} |\tilde{t}_{\pi\gamma}^{NN}|
  \vec{d}=-2\vec{k},m_{d}\rangle & = & \pi\sqrt{2E_{d}}\,
  u_{0}(0)\sum_{m_{1},m^{\prime}_{1},m^{\prime}} \langle -\vec{q},
  m_{1}^{\prime} | t_{\pi\gamma} | -\vec{k}, m_{1} \rangle
  \nonumber \\ & & \times ( 1/2 \, m_{1}^{\prime}, 1/2 \,
  m^{\prime}|1\, m_{s} )\,\, ( 1/2 \, m_{1}, 1/2 \,
  m^{\prime} |1 \, m_{d}) \, .
\end{eqnarray}
Performing the spin summation in (\ref{c1}) explicitly one finds
\begin{equation}\label{c3}
  \frac{1}{6} \sum_{m_{d}} \sum_{s m_{s}}
  |\tilde{t}_{\pi\gamma}^{NN}|^{2} = \pi^{2} u_{0}^{2}(0) E_{d}
  \sum_{m_{1} m_{1}^{\prime}} |t_{\pi\gamma}|^{2} \, ,
\end{equation}
where $m_{1}$ and $m_{1}^{\prime}$ are the spin projections of the
nucleon in the initial and final states, respectively. Due to the
small binding energy of the deuteron we can use the approximation
\begin{equation}
  \frac{2 \sqrt{M_{N}^{2}+\vec{k}^{\, 2}}}{\sqrt{M_{d}^{2}+
      (2\vec{k})^{2}}} \sim 1 \, .
\end{equation}
In the c.m. frame, the invariant mass of the $\pi$N system is given by
\begin{equation}
  W_{\pi N} = \omega_{\gamma} + E_{\vec{k}} = \omega_{\vec{q}} +
  E_{\vec{q}} \, .
\end{equation}
From (\ref{c1}) with the help of (\ref{c3})
it is now easy to verify equation (\ref{g19})
\begin{eqnarray}
  \frac{d^{5}\sigma}{d^{3}P_{\pi N}d\Omega_{\pi}} & = & \frac{1}{64
    \pi^{2}} \frac{|\vec{q}|M_{N}^{2}}{\omega_{\gamma}W_{\pi N}^{2}}
  \frac{u_{0}^{2}(0)} {4\pi} \sum_{m_{i} m_{f}} |t_{\pi\gamma}|^{2} \\ 
  & = & \frac{u_{0}^{2}(0)}{4\pi} \frac{d\sigma_{\pi\gamma}}{d\Omega_{\pi}} 
  \, .  \nonumber
\end{eqnarray}
Here a second approximation is used implicitly because we compare the
elementary and the quasifree process for a given invariant mass of the
$\pi$N system. Due to the binding of the nucleons in the deuteron
there is a small difference in the corresponding photon energies
$\omega_{\gamma}^{q.f.}$ and $\omega_{\gamma}^{el.}$. Expanding
the ratio of the photon energies for a given
$W_{\pi N}$
with respect to 
the relative binding energy $\epsilon = (2M_{N} -
M_{d})/M_{N} \sim 0.0024$  yields
\begin{eqnarray}
  \frac{\omega^{q.f.}} { \omega^{el.}} = 1 + \frac{ {M_{N}^4} +
      6\,{M_{N}^2}\,{W_{\pi N}^2} + {W_{\pi N}^4} } {4\,{W_{\pi
          N}^2}\,\left( {W_{\pi N}^2} -{M_{N}^2} \right) } \, \epsilon +
    {{{\rm O}(\epsilon )}^2}\, . 
\end{eqnarray}
The coefficient of the linear term is about $6.3$ at pion threshold
and decreases with increasing $W_{\pi N}$. Thus this approximation is
very well justified.

\newpage

\newpage

\newpage

\newpage

\newpage

\newpage
\end{document}